\documentclass{aa}
\usepackage{graphicx}
\usepackage[authoryear]{natbib}
\bibpunct{(}{)}{;}{a}{}{,}
\usepackage{txfonts}

\newcommand{\ca}{\ion{Ca}{ii} K}

\newcommand{\gsim}{\lower.4ex\hbox{$\;\buildrel >\over{\scriptstyle\sim}\;$}}

\def\apjl{ApJL}

\begin{document}

\title{The relationship between chromospheric\\ emissions and magnetic field strength}
  \author{M. Loukitcheva
           \inst{1,2}
           \and
           S. K. Solanki
           \inst{1}
           \and
           S. M. White
           \inst{3}
            }
  \offprints{M. Loukitcheva}

   \institute{Max-Planck-Institut f\"ur Sonnensystemforschung, D-37191 Katlenburg-Lindau,
   Germany\\
              \email{lukicheva@mps.mpg.de}
          \and
              Astronomical Institute, St. Petersburg University, 198504 St. Petersburg, Russia
          \and
              Astronomy Department, University of Maryland, College Park, MD 20742
              }

\date{Received  / Accepted }

\titlerunning{Chromospheric structure from multi-band observations}
\authorrunning{Loukitcheva et al.}

\abstract{}{We analyze observational data from 4 instruments to
study the correlations between chromospheric emission, spanning the
heights from the temperature minimum region to the middle
chromosphere, and photospheric magnetic field.} {The data consist of
radio images at 3.5~mm from the Berkeley--Illinois--Maryland Array
(BIMA), UV images at 1600~\AA\ from TRACE, \ca-line filtergrams from
BBSO, and MDI/SOHO longitudinal photospheric magnetograms. For the
first time interferometric millimeter data with the highest
currently available resolution are included in such an analysis. We
determine various parameters of the intensity maps and correlate the
intensities with each other and with the magnetic field.} {The
chromospheric diagnostics studied here show a pronounced similarity
in their brightness structures and map out the underlying photospheric
magnetic field relatively well. We find a power law to be a good
representation of the relationship between photospheric magnetic
field and emission from chromospheric diagnostics at all wavelengths.
The dependence of chromospheric brightness on magnetic field is
found to be different for network and internetwork regions.}{}


\maketitle

\section{INTRODUCTION}

The solar chromosphere is known to be highly structured at different spatial
scales and the range of chromospheric inhomogeneities is best
revealed when observed at different wavelengths from the extreme
ultraviolet to millimeter waves \citep[see,
e.g.,][]{2004IAUS..223..195S}. Even far from activity
complexes, in the quiet Sun magnetic field is generally believed to be
responsible for chromospheric structuring. This is consistent with the strong
correlation found between excess
emission in the cores of the \ca\ and H resonance lines (the primary
chromospheric diagnostic) and the presence and strength of quiet-Sun
magnetic flux \citep[e.g.][]{1959ApJ...130..366L,1989ApJ...337..964S}.
Consequently, these resonance lines commonly serve as indicators of
changes in the chromospheric structure related to global magnetic
activity and heating of the outer atmosphere of the Sun and cool
stars \citep{1991SoPh..134...15R,1989ApJ...337..964S}. In addition to
the chromospheric emission related to the magnetic field,
\citet{1987A&A...172..111S,1992A&A...258..507S} has suggested that
part of the radiative flux from solar and stellar chromospheres,
referred to as basal flux, has a nonmagnetic origin and is believed
to stem from acoustic heating of the chromosphere \citep[see,
e.g.][]{1991SoPh..134...15R}.

The primary brightness patterns revealed in two-dimensional
images obtained in the ionized calcium H and K lines include bright
plages, chromospheric network located at the boundaries of
supergranular cells, and bright points in the interior of the cells
-- internetwork  grains or \textit{K2V} bright points. Chromospheric
network features are known to display a one-to-one spatial
correspondence with regions of enhanced photospheric magnetic field
\citep{1975ApJ...200..747S,1982SoPh...80..227S,1998SoPh..179..253N,
2000A&A...363..279S}, while the (magnetic) origin of the
internetwork grains is
a controversial issue.
From the analysis of simultaneous Kitt Peak \ca-line
spectroheliograms and photospheric magnetic area scans,
\citet{2000A&A...363..279S} showed the association of internetwork
\textit{K2V} bright points with magnetic features, confirming the
earlier results of \citet{1982SoPh...80..227S}. However, from
analysis of \ca-line and magnetic field data from the BBSO,
\citet{1998SoPh..179..253N} have found that not all internetwork
bright points are associated with magnetic fields. The existence of
2 distinct classes of cell-interior grains, which differ in temporal
behaviour and spatial motion and have a different origin, was revealed
by observations with the Swedish Solar Telescope reported by
\citet{1992ASPC...26..161B}, whereas \citet{1999ApJ...517.1013L}
found no obvious relation of the internetwork brightenings to the
magnetic field, and \citet{1996A&A...316..196R} concluded that they
are a totally nonmagnetic phenomenon.

Quantitative studies of the relationship between magnetic field and the chromosphere
(represented by the calcium K and H-lines), in both the active and quiet
solar atmosphere, have been carried out by a number of authors
\citep[e.g.][]{1975ApJ...200..747S,1989ApJ...337..964S,1998SoPh..179..253N,1999ApJ...515..812H,2005MmSAI..76.1018O,2007A&A...466.1131R}.
In the quiet-Sun chromosphere \citet{1975ApJ...200..747S} and
\citet{1998SoPh..179..253N} found a linear correlation
between the K-line intensities and the absolute values of the
magnetic field strength. \citet{1989ApJ...337..964S} extended
this relationship to solar active regions and suggested
that the relation between the K-line excess flux density (after the
subtraction of the basal flux from the observed \ca-line flux) and
the magnetic field flux density is best described by a power law
with an exponent of about 0.6. Considering 4 types of brightness
structures, which included active regions, decaying active regions,
the enhanced network and quiet (weak) network,
\citet{1999ApJ...515..812H} ascertained that in both active and
quiet Sun the relation between the magnetic flux density and calcium
residual intensity follows a power law with an exponent of 0.5.
\citet{2005MmSAI..76.1018O} cited a power-law exponent value of 0.66
for the quiet Sun, while \citet{2007A&A...466.1131R} obtained a
lower exponent of 0.2 for locations in a
quiet-Sun region, and higher values of 0.4--0.5 for network
locations, depending on the magnetic field threshold value.

\citet{1991A&A...252..203R} found power-law relations between
excess flux densities in stellar chromospheric lines with differing
temperature responses, with the exponents increasing with increasing
difference between their temperatures of formation. Relations
between fluxes originating in the same atmospheric regime were found
to be close to linear (i.e., having a power-law exponent of unity). For the quiet
Sun the high degree of co-spatiality of the brightness features seen
in the \ca\ line and the \textit{TRACE} UV channels has been
demonstrated by \citet{1999ASPC..183..383R}.

Almost all these studies have been limited to the visible and UV
spectral range. An exception is the work of
\citet{1991ApJ...383..443L}, who studied the cross-correlation
statistics of the brightness variations in the submm (850$\mu$m) and
\ion{Ca}{ii} K-line images. Their results are confined to the
temperature minimum region and the heights of the low chromosphere.
No detailed studies have compared millimeter emission, which
originates from the middle chromosphere, with other chromospheric
indicators. From observations at
350, 850 and 1200 $\mu$m with the James Clerk Maxwell Telescope
at a resolution of about 20\arcsec,
\citet{1995ApJ...453..511L} reported submillimeter chromospheric
brightness structures which
resemble the supergranular network, although the resolution was
marginal for such a study. Initial qualitative results of the
comparison of chromospheric structure, identified in millimeter
images obtained with BIMA at 3.5 mm with a resolution of about
10\arcsec, with calcium images from BBSO and
line-of-sight photospheric magnetograms from SOHO/MDI were given in
\citet{2006A&A...456..697W}.

In this paper we present an observational study of the
relationships between the photospheric magnetic field and emission
at different temperatures from chromospheric heights. We study the
chromospheric observational data referring to the temperature
minimum region (UV continuum at 1600~\AA),  low chromosphere (\ca\
line) and middle chromosphere (continuum at 3.5 mm). This is the
first analysis of this type that includes millimeter observational
data. Unfortunately, simultaneously recorded observational data on
the chromospheric magnetic field suitable for such an analysis are
unavailable. For this reason, we focus on the MDI/SOHO longitudinal
photospheric magnetograms and study the dependence of
the chromospheric intensity at different wavelengths on the
photospheric magnetic field strength.

In Sec. 2 and 3 we describe the data and instruments used for the
analysis and provide the details of the data reduction. In Sect.~4
the topography of the quiet chromosphere, chromospheric intensity
distributions and cross-correlations between the magnetic field and
chromospheric emission are presented. In Sect.~5 we summarize and
discuss the results.

\section{OBSERVATIONAL DATA}\label{section1}

We analyze observations from the
Berkeley--Illinois--Maryland Array (BIMA) at 85 GHz ($\lambda\,=\,3.5$ mm),
\ca-line filtergrams from
the Big Bear Solar Observatory (BBSO), UV images at 1600~\AA\ from
the Transition Region and Coronal Explorer
\citep[TRACE,][]{1999SoPh..187..229H} and magnetograms from the
Michelson Doppler Imager \citep[MDI,][]{1995SoPh..162..129S} on
the Solar and Heliospheric Observatory (SOHO) satellite, obtained on
May 18, 2004. The data consist of temporally resolved images of the
measured intensities, which are given in Kelvin for millimeter data
(see Sec.~\ref{sec2.4}), in relative units for \ca\--line and
1600~\AA\ data (Sec.~\ref{sec2.2} and \ref{sec2.3}, respectively),
and in Gauss for MDI photospheric magnetograms (Sec.~\ref{sec2.1}).
The overview of the data (including instruments, operating
wavelengths, spatial resolution, cadence and image pixel size) used
in the present study is given in Table~\ref{table1}.

\begin{table*}
  \centering
  \caption{Overview of data taken on May 18, 2004}\label{table1}
    \begin{tabular}{ccccccc}
            \hline\hline
            \noalign{\smallskip}
            Instrument & Passband & $\lambda$ & $\Delta\lambda$(\AA) & Spatial resolution (\arcsec) & Time cadence & Pixel size (\arcsec)\\
            \noalign{\smallskip}
            \hline
            \noalign{\smallskip}
            MDI/SOHO & \ion{Ni}{I} & 6768\AA & 0.094& 4 & 1~min & 1.96 \\
            TRACE & \ion{C}{IV}+cont & 1600\AA & 275 & 1 & 30~min & 0.5\\
            BBSO & \ca\ & 3933\AA & 0.6 & 2-4 & 30~s & 0.645 \\
            BIMA & continuum & 3.5 mm & & 12 & 10~s & 3.0 \\

            \noalign{\smallskip}
            \hline
    \end{tabular}
\end{table*}

\begin{figure*}
   \centering
   \includegraphics[width=0.75\textwidth,angle=90]{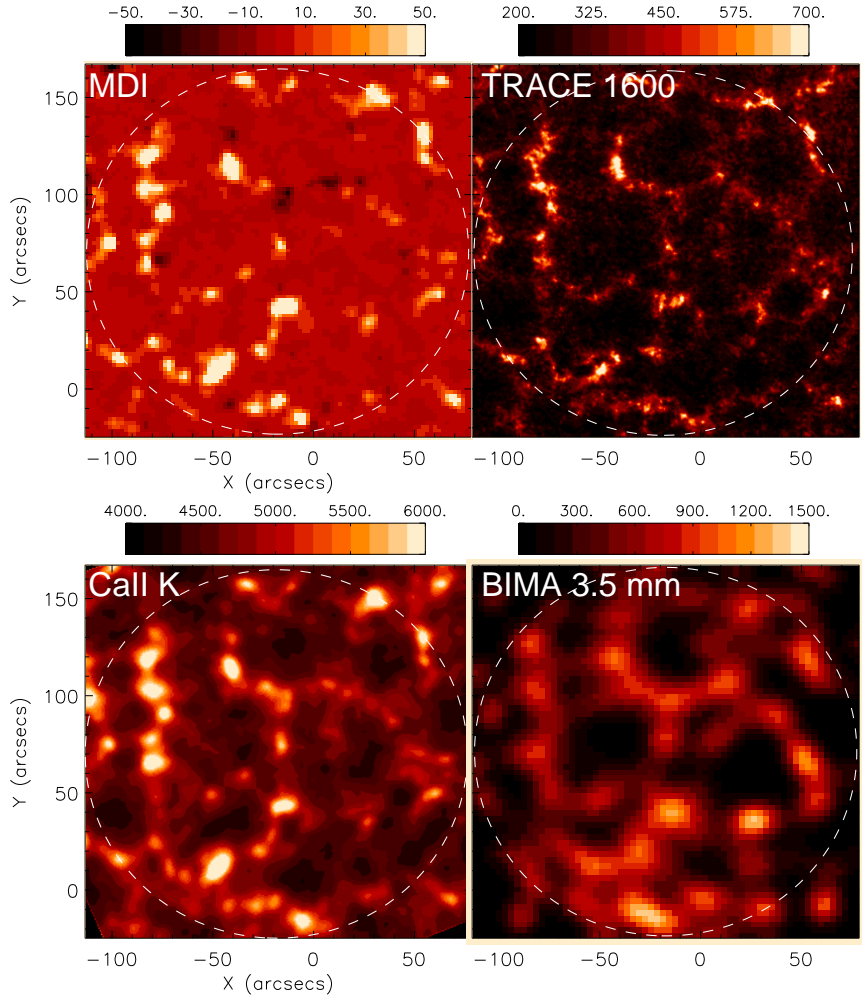}
   \caption{Portrait of the solar chromosphere averaged over 3.5~h at the center of the
Sun's disk at 4 different wavelengths on May 18, 2004. From top left
to bottom right: MDI longitudinal photospheric magnetogram
(resolution of 4\arcsec), TRACE 1600~\AA\ image (resolution of
1\arcsec), \ca-line center image from BBSO (resolution of
2-4\arcsec) and BIMA image at 3.5 mm (resolution of 12\arcsec). The
images are created by averaging over the 3.5h period of the BIMA observation.
The details of the data units are given in
Sect.~\ref{section1}. Dashed circles mark the 96\arcsec\ radius BIMA FOV
where we regard fluxes to be reliable. \textit{X} and \textit{Y} axes are in
arcsec from the disk center, which is located at $(0,0)$.}
              \label{fig1}%
    \end{figure*}

   \begin{figure*}
   \centering
   \includegraphics[width=0.75\textwidth,angle=90]{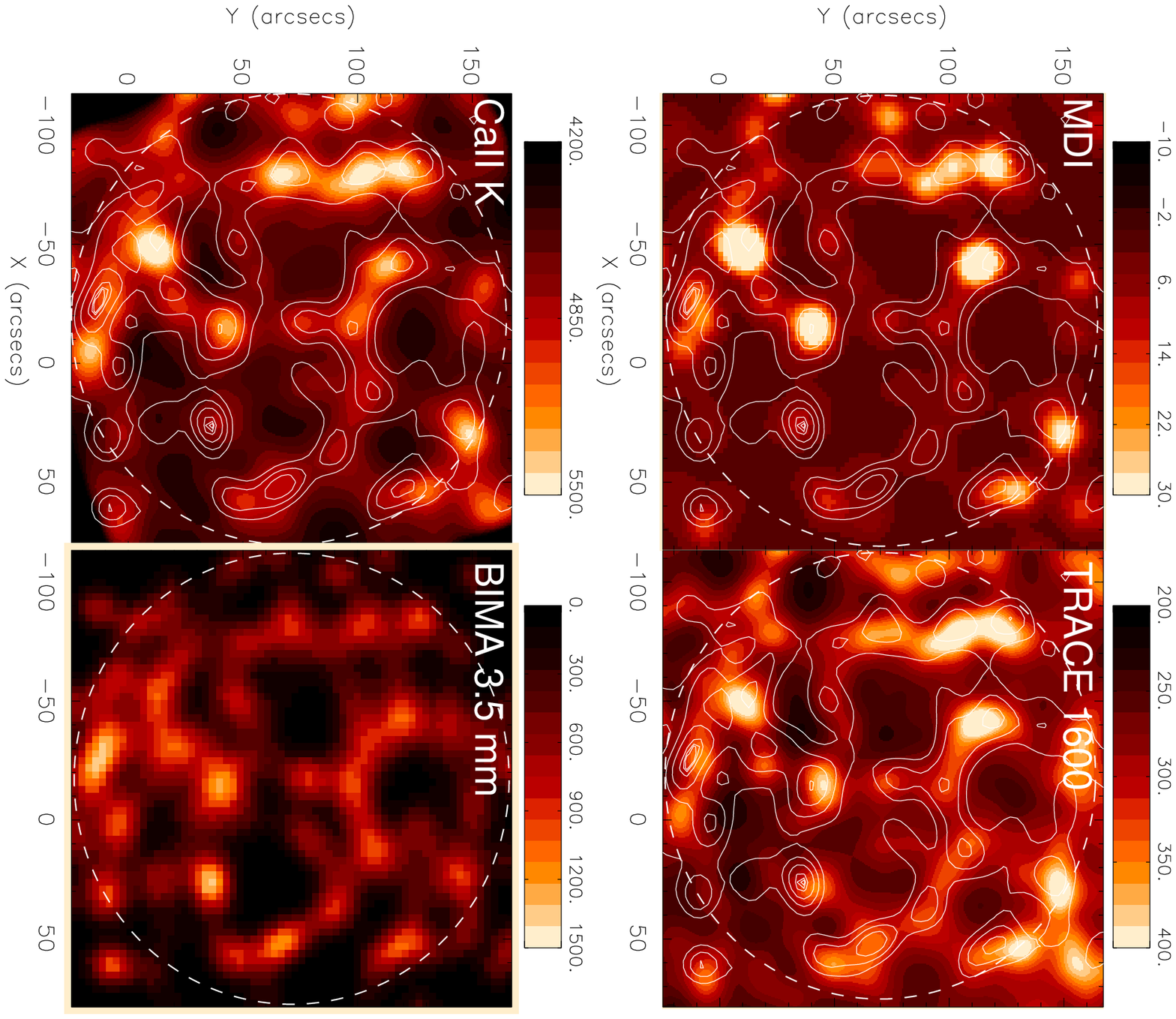}
   \caption{The same as Fig.~\ref{fig1} plotted at the BIMA resolution of
   12\arcsec. In the top left panel the absolute value of the MDI magnetogram is depicted.
   Overlays of contours of 3.5~mm emission at (30, 50, 70, 90, 95, 98)\% of maximum brightness are
   overplotted in white.}
              \label{fig2}%
    \end{figure*}

\subsection{BIMA observations at 3.5 mm}\label{sec2.4}

The 3.5~mm observations of a quiet-Sun area near the center of the
disk were acquired with 9 BIMA antennae during the time interval
16:43--20:21~UT ($\sim$3.5 hours) when atmospheric phase conditions
were good. The data were taken at a cadence of 10 seconds. The
interferometer was in its most compact (``D'') configuration for
these observations, providing a spatial resolution of order
12\arcsec\ within a 2\arcmin\ field of view (FOV). The images were
deconvolved using a maximum entropy method (MEM) and restored with a
Gaussian beam. The details of the interferometric image restoration
of the BIMA data, including description of the limitations imposed
by the MEM deconvolution, are given in \citet{2006A&A...456..697W}.
The final images were 128 pixels square, with a cell size of
3\arcsec\ and the flux in the inner 80 pixels of the FOV with the
center at (-17\arcsec, 71\arcsec) in heliocentric coordinates. All
images were corrected for the primary beam
response and flux units were converted to brightness temperatures
for a 12\arcsec\ beam. Data within a 96\arcsec\ radius of the pointing center
are used in the analysis below. The zero level in the BIMA image corresponds
the coldest feature on the surface in the chromosphere that is
optically thick at 3.5~mm wavelength. This should be at a
temperature of order 6500--7000 K. The temperature fluctuations in
the BIMA image represent direct temperature measurements above this minimum value
on the optically thick surface in the chromosphere, ranging from 0 to 1100
K in the map for the entire 3.5 hour period.

\subsection{\ca\ line observations at 393.3~nm}\label{sec2.2}

The \ca-line observations were obtained with the 25-cm refractor of
BBSO. A 5h sequence of filtergrams in the core of the Ca \rm{II}
K-line (393.3~nm) with a bandpass of 0.6~\AA\ was recorded by a
KX-260 CCD camera, which has $512 \times 512$ pixels and a dynamic
range of 14 bits. The exposure time was 250~msec, the cadence was
30~sec, recorded field of view was 300\arcsec $\times$ 300\arcsec\
at a pixel size of 0.645\arcsec. The observations were taken under
moderate to fair seeing conditions. The K-line data were
preprocessed, which included calibration by dark and flat-field
frames \citep{2001SoPh..202...63D}. The intensity scale of the
images was equalized by setting the quiet-Sun integrated background
intensity in all images to the same value. No absolute flux
calibration was available for the data, hereafter we use the
\ca-line intensities in arbitrary units. We note that the bandwidth
of the calcium filter tuned to the line center was too large to
separate the K-line peaks and thus the images obtained depict the
structure of the solar chromosphere collected from a wide range of
heights, but dominated by mid-chromospheric contributions, as sampled
by the $K2$ peaks.

\subsection{TRACE observations at 1600 \AA}\label{sec2.3}

In parallel with the BIMA observations a total of 14 TRACE images
with a cadence of 20 minutes
were recorded at a wavelength of 1600~\AA\ at full resolution
(0.5\arcsec\ pixel size). The images were preprocessed, which
included subtraction of the dark pedestal and current, as well as
division by a corrected flat field. The data were normalized to the
exposure time, giving data numbers per second
($DN s^{-1}$) with no absolute calibration.
Both standard models of the quiet-Sun atmosphere such as FAL C
\citep{1993ApJ...406..319F} and non-LTE radiation hydrodynamic
simulations such as those of \citet{2005ApJ...625..556F} place the
disk-center radiation at 1600~\AA\ in the temperature minimum region
at a height of around 500~km above the photosphere
\citep{1999SoPh..187..229H,2001A&A...379.1052K,2005ApJ...625..556F},
although a contribution from $10^5$~K transition-region gas is also
present due to the \ion{C}{IV} lines. In the quiet Sun this
contribution is expected to be small
\citep{1998SoPh..183...29H,2001A&A...379.1052K}.

\subsection{MDI magnetograms}\label{sec2.1}

One-minute cadence full-disk MDI magnetograms \footnote{lev1.8 data.} \citep[spatial resolution
of 4\arcsec,][]{1995SoPh..162..129S} were selected for the 3.5~h
BIMA observing period, which resulted in 188 images of
the proxies of the photospheric magnetic field. The noise in the
original 1-min MDI magnetograms is typically about 20 Gauss
\citep{1995SoPh..162..129S}, which is comparable to the magnetic
signal of quiet-Sun network features. By averaging in time over 24
consecutive individual magnetograms (30-min sequence) we reduced the
noise to 4 Gauss. Averaging over the whole 3.5~h observing period
(188 images) resulted in a level of the random noise of about 1.5
Gauss. Prior to averaging the influence of solar rotation on the
time series was removed (see Sect.~\ref{section2}).

\section{DATA COALIGNMENT AND REDUCTION}\label{section2}

We formed 4 observational data cubes $I(x,y,t)$, which were
subsequently aligned using a two-step procedure. The first step
involved correction of differential solar rotation by using standard
procedures of SolarSoftware in the IDL environment. All images were
rotated to the fiducial time of 20:00~UT. In the second step the
shift of the images was determined and applied by a 2D
cross-correlation of each image of the sequence to the first image
of the sequence.

After a sub-arcsecond accuracy was reached in the alignment of the
data cubes we averaged the images in time over the whole observing
period (16:43--20:20~UT) to obtain the mean (time-averaged) MDI
magnetogram, \ca~ filtergram, TRACE 1600~\AA\ image and BIMA image
at 3.5 mm. Additionally we averaged over the uninterrupted sequence
of images of good quality within the time interval 19:00--19:30~UT
to check the stability of the large-scale quiet-Sun structures. We
based our correlation analysis on the mean images to avoid the
scatter introduced by the chromospheric oscillations within cell
interiors and the chromospheric network. Consequently, by averaging
the images we also suppress proper motions of the K-grains and of
bright points in the network.

Finally, the full disk $B_{\|}$ MDI images were chosen as the
coordinate reference against which all other mean images were
co-aligned by using cross-correlation methods. Absolute values of
the MDI magnetograms were formed prior to alignment. The remaining
image displacements do not exceed 1\arcsec\ and we believe that they
do not affect the results of our analysis. A 3.5h time-averaged
image of the BIMA FOV made at each of the 4 wavelength bands is
shown in Fig.~\ref{fig1} at the original spatial resolution of each
band.

The cross-correlation analysis between the chromospheric images and
photospheric magnetograms was done both for the MDI magnetogram
resolution (4\arcsec) and for the resolution of the BIMA data
(12\arcsec). In the former case the \ca-line and TRACE 1600~\AA\
images were rebinned to make the size of the pixels equal to that of
the magnetograms. In the latter case we degraded all images to the
resolution of the BIMA data prior to cross-correlating them. The
shift between the BIMA images and the MDI magnetograms did not
exceed 1 BIMA image pixel ($<3$\arcsec). Figure~\ref{fig2} shows
example time-averaged images at BIMA resolution.

\section{RESULTS}{\label{section3}}

\subsection{Mapping of chromospheric structure}

Figure~\ref{fig1} depicts the time-averaged quiet solar chromosphere
at 3 different heights from the temperature minimum region (TRACE
1600~\AA\ image) through the lower chromosphere (\ca-line
filtergram) to the middle chromosphere (BIMA image at 3.5 mm) at the
original resolution of the instruments (see Table~\ref{table1}) as
well as the corresponding MDI/SOHO line-of-sight photospheric
magnetogram. Since the QS region being analyzed lies near
the disk center, the line-of-sight magnetic field, measured by MDI,
can accurately be regarded as the vertical component of the magnetic
field vector. Of particular interest is the relation of the BIMA
image to the others. Visual inspection of Fig.~\ref{fig1} reveals
strong morphological similarities between different chromospheric
emissions and the photospheric magnetic field, most prominently on
spatial scales of the enhanced magnetic flux and chromospheric
network (brightenings in all 4 frames). In snapshot images (not
plotted) brightenings appear not only at supergranular cell
boundaries, but also within the cells. Averaged in time, bright
network becomes more pronounced, while low-intensity short-lived
internetwork bright points are significantly suppressed and eventually
vanish with increased averaging. In general, use of time averages
reduces differences between the 4 images due to the brightness
variability of the structures.

The impression of a high degree of correlation of photospheric
magnetograms (absolute value) with the \ca-line, 1600~\AA\ and
3.5~mm images is confirmed quantitatively. Within a radius of
96\arcsec\, the linear correlation coefficient between the absolute
magnetogram and calcium image is 0.75, while it is 0.63 between the
magnetogram and emission at
1600~\AA.

The degree of spatial agreement between the radio images and the other
chromospheric diagnostics and the photospheric magnetic field is more
clearly visible in maps degraded to the BIMA resolution of
12\arcsec, shown in Fig.~\ref{fig2} with radio contours overlaid on the other images at
(30, 50, 70, 90, 95, 98)\% of maximum mm brightness. The correlation
coefficients between the mm image and the magnetogram are 0.56
(linear) and 0.72 (rank), while for the mm image and \ca-line image
they are 0.65 and 0.69, respectively. At a resolution of 12\arcsec\
the magnetogram shows a higher correlation with the calcium and
1600~\AA\ images than at 4\arcsec\ resolution, with the linear
correlation coefficient being 0.82 and 0.74, respectively.

  \begin{figure}[!htb]
   \centering
   \includegraphics[width=0.4\textwidth]{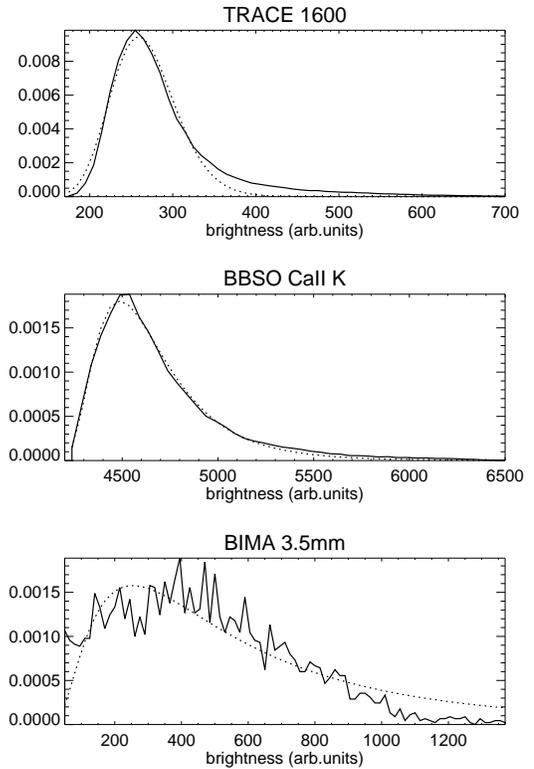}
   \caption{Intensity distribution histograms over the 2-D quiet-Sun region near the disk center
   obtained, \textit{from top to bottom}, in the TRACE 1600 \AA\ channel,
   the \ca\ line and at 3.5 mm. In all three panels dotted curves depict log-normal fits.}
              \label{fig}%
    \end{figure}

\subsection{Histograms of chromospheric intensity distribution}

With sufficiently good statistics the brightness histogram is
expected to have a log-normal shape, comparable with a single
population \citep{2000A&A...362..737P}. To test whether this is the
case for the chromospheric intensities studied here we built
intensity distribution histograms (observed frequency distribution
normalized to unit area) for the selected 2D images at 1600 \AA\ and
3.5 mm and fitted them with a single log-normal distribution
function. Following \citet{2007A&A...468..695F} histograms of the
\ca\ intensity were fitted with a modified log-normal function
having an additional parameter -- an offset which represents line
scattering of the photospheric background radiation. In
Fig.~\ref{fig} we show the derived histograms for the observed
intensities at 1600 \AA, in the \ca\ line and at 3.5 mm and their
log-normal fits with solid and dotted curves respectively.

A log-normal with an offset provides a good fit to the observed \ca\ line
intensity distribution, although the observed tail enhancement is
slightly higher than that of the fitted log-normal (see central
panel in Fig.\ref{fig}).
For the 1600~\AA\ intensities the quality of the fit
diminishes at low and high flux values and the
departure from a log-normal form increases. Finally, due to a
substantially higher noise level in the millimeter brightness data
(and the smaller number of resolution elements available for the histogram)
it is difficult to evaluate the quality of the log-normal fit for
this dataset. On the whole, a tendency for bigger departures from a
log-normal can be observed when spatial resolution is decreased.
However, we cannot rule out the possibility that we are seeing the effects
of different heating mechanisms in cell interiors and network lanes being
manifested in the intensity distributions.

 \begin{figure*}
   \centering
    \includegraphics[width=0.55\textwidth,angle=90]{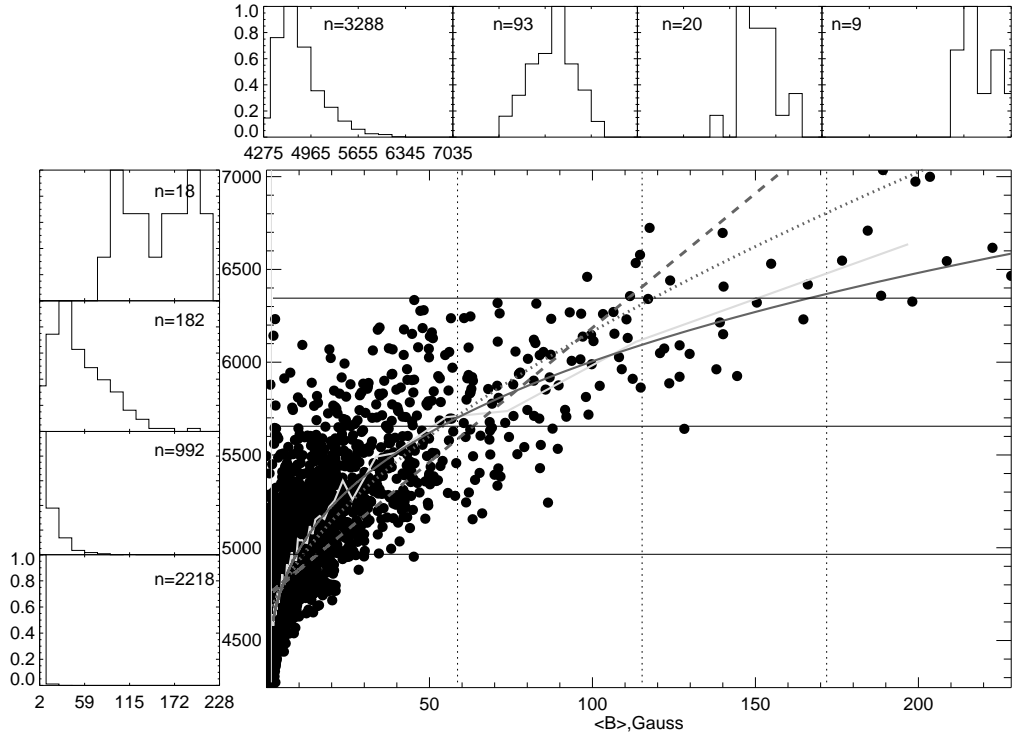}
   \caption{The \ca-line intensity vs. the absolute value of the magnetic flux density for a 2\arcsec\ pixel
   size. The images are averaged in time over 3.5h. The vertical line at 1.5~G in the central panel represents the 1$\sigma$ noise limit of the magnetogram signal.
   The fit results are represented by a solid curve for a power-law fit with free exponent, dotted curve for a
   power-law fit with the exponent fixed at 0.6 and dashed for a linear fit. The light grey solid curve outlines the
   \ca\ intensity binned in 70 intervals containing an equal number of points.
   Histograms on the left show the distribution of magnetic field strength in 4 subranges of the \ca-line flux.
   Histograms on top of the central panel depict the distribution of calcium intensity in 4 subranges of the magnetic
   flux. For each histogram the corresponding range of values considered is marked in the main figure: by horizontal solid
   lines for magnetic flux distributions and by vertical dotted lines for the \ca-line flux
   histograms. The numbers of counts, \textit{n}, represented in the histograms are indicated.}
              \label{fig3}%
    \end{figure*}

\begin{figure*}
   \centering
  \includegraphics[width=0.55\textwidth,angle=90]{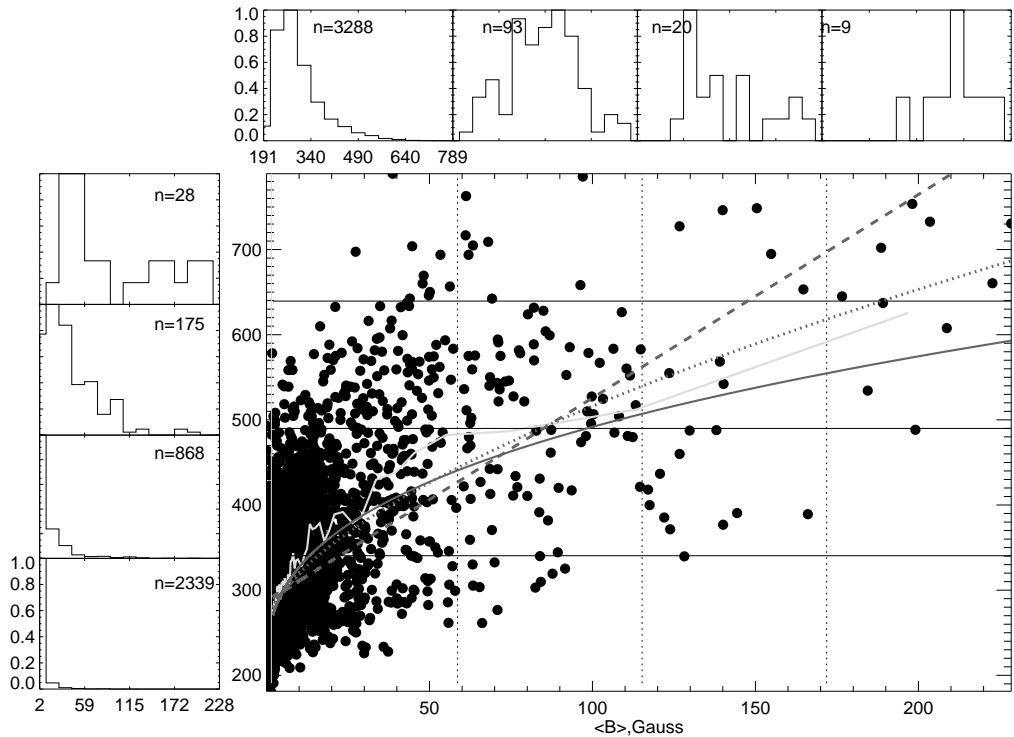}
   \caption{The same as Fig.~\ref{fig3} for the TRACE 1600~\AA\ flux at the 4\arcsec\ resolution of MDI.}
              \label{fig4}%
    \end{figure*}

\begin{figure*}
   \centering
   \includegraphics[width=0.55\textwidth,angle=90]{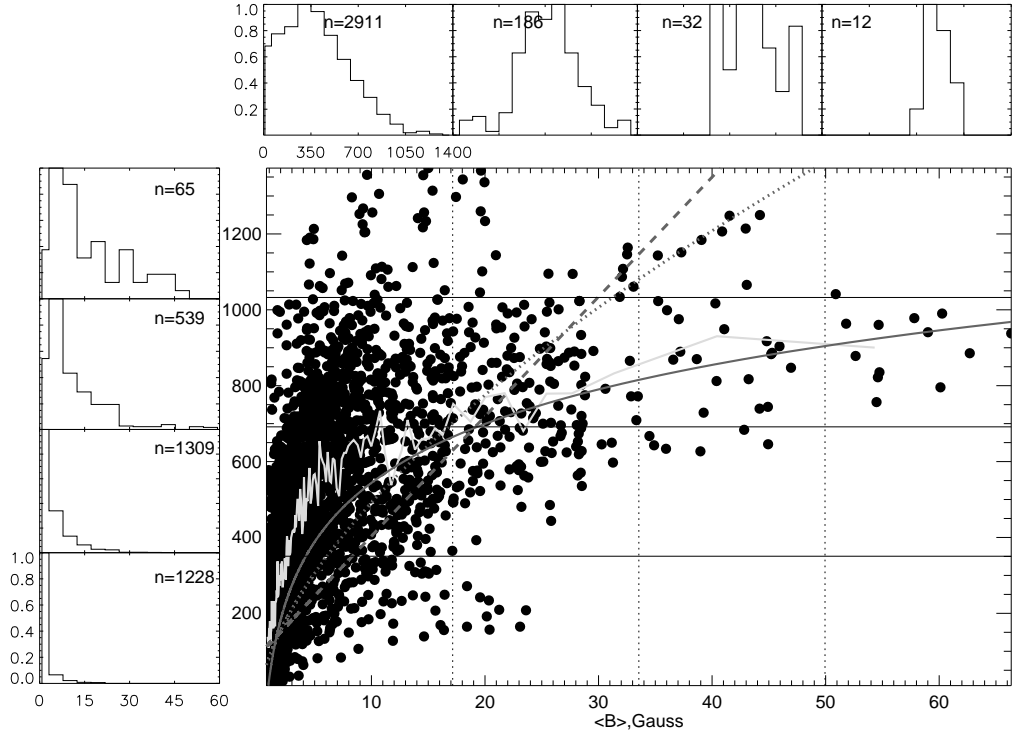}
   \caption{The same as Fig.~\ref{fig3} for the brightness temperature at 3.5~mm at the BIMA resolution of 12\arcsec.}
              \label{fig5}%
    \end{figure*}

\subsection{Relation between chromospheric emission and photospheric magnetic field}

In order to study the relationship between the intensity of
chromospheric bright features and the intensity of the associated
magnetic elements, we computed scatter plots for the pairs of
images. Firstly, we considered the pointwise (pixel-to-pixel)
correlation globally in the field of view without regard to the
location of the points. In Fig.~\ref{fig3} we show the resulting
cross-correlation statistics of the brightness features in the
time-averaged calcium image and the photospheric longitudinal
magnetogram. The pixel intensity correlation is plotted in the main
frame of the figure, with histograms of the calcium intensity in 4 subranges of
the magnetic flux density superposed in the upper frame and of the
magnetic field in 4 subranges of the calcium intensity to the right.
We employed averaging over time to reduce the noise level in the
scatter plots.

We searched for a quantitative relation between the quiet-Sun mean
magnetic flux density (its absolute value) per pixel of 2\arcsec\
and \ca-line intensity, looking for the best least-squares fit among
functions of 2 types: linear \citep[i.e. a power law with
unit exponent, Eq.~\ref{1},
following][]{1975ApJ...200..747S,1998SoPh..179..253N} and a
``generalized'' power law \citep[Eq.~\ref{2},
following][]{1989ApJ...337..964S}.

    \begin{equation}\label{1}
      I=I_{0}+a\cdot|B|
   \end{equation}

\begin{equation}\label{2}
    I=I_{0}+a\cdot|B|^{b}
\end{equation}

The item $I_{0}$ in Eqs.~\ref{1} and  \ref{2} denotes ``zero'' or
``basal'' flux, introduced by \citet{1987A&A...172..111S}. For
comparison with the results of \citet{1989ApJ...337..964S} and of
others, who obtained a power law exponent of about $0.6$ for active
and quiet regions, we also calculated the best fit with the
power-law exponent, $b$, fixed at $0.6$. Fits to the data points and
to binned values (averaged over all pixels within a narrow range of
magnetic flux values) points (see central panel in Fig.~\ref{fig3})
gave very similar results. Only data points with an absolute value
of the magnetic flux exceeding the $1\sigma$-noise level were
considered for fitting.

In the same manner we studied the relationship between the observed
UV intensity at 1600~\AA\ and absolute magnetic flux density per
2\arcsec\ aperture, which is depicted in Fig.~\ref{fig4}, and
between intensity at 3.5~mm and absolute magnetic flux density per
12\arcsec\ aperture, plotted in Fig.~\ref{fig5}. The decrease in
range of $<B>$ is due to the lower spatial resolution of mm data. In
Table~\ref{table2} we list the results of the function fitting to
the data points for all 3 relationships investigated, and in particular we
compare $\chi^2$ values for 3 different fitting functions obtained
for Poisson weighting and list the power law exponents obtained.

\begin{table*}
  \centering
  \caption{Least-squares deviation, $\chi^2$, of the best fits to the relationships between the photospheric magnetic field and chromospheric emissions}\label{table2}
    \begin{tabular}{c|c|c|cr}
            \hline\hline
            \noalign{\smallskip}
            Flux-flux & Linear$^{\mathrm{a}}$ & Power law$^{\mathrm{b}}$ with $b=0.6$ & \multicolumn{2}{c|}{Power law$^{\mathrm{b}}$ } \\
            \cline{2-5}
            \noalign{\smallskip}
            & $\chi^2$ & $\chi^2$ & $\chi^2$ & $b$ \\
            \hline
            \noalign{\smallskip}
            $|B| - I_{CaII K}$ & 11.97 & 9.17 & 8.44 & 0.31 $\pm$ 0.01   \\
             \noalign{\smallskip}
            $|B| - I_{1600 \AA}$ & 11.53 & 10.64 & 10.47 & 0.36 $\pm$ 0.01  \\
             \noalign{\smallskip}
            $|B| - I_{3.5 mm}$ & 161.68 & 132.78 & 111.06 & 0.20$^{\mathrm{c}}$ $\pm$ 0.01  \\
            \noalign{\smallskip}
            \hline
    \end{tabular}
\begin{list}{}{}
\item[$^{\mathrm{a}}$] see Eq.~\ref{1}
\item[$^{\mathrm{b}}$] see Eq.~\ref{2}
\item[$^{\mathrm{c}}$] calculated for $|B| \geq 4$~G
\end{list}

\end{table*}

As seen from the scatter plots in Figs.~\ref{fig3}, \ref{fig4} and
\ref{fig5} all three quantities, \ca-line, UV at 1600~\AA\ and
3.5~mm emission, show a prominent non-linear dependence on
photospheric magnetogram signal.
This conclusion is confirmed by the best-fit curves plotted in
Figs.~\ref{fig3}, \ref{fig4} and \ref{fig5} and by the $\chi^2$
values listed in Table~\ref{table2}, which reach the smallest values
for the fit in the form of a power law. The best-fit value of the
power-law exponent is $0.31$ for the relationship between the
calcium intensity and the magnetic flux density, which lies between
the high value of $0.6 \pm 0.1$ and $0.66$ found by
\citet{1989ApJ...337..964S} and \citet{2005MmSAI..76.1018O} and the
lower value of $0.2$ obtained by \citet{2007A&A...466.1131R}. For
1600~\AA\ intensity vs. magnetic flux we obtain an exponent of
$0.36$. These results do not change if we use, e.g., 10-minute or
30-minute averages instead of the 3.5h average used above. However,
the scatter of the points at a given $<B>$ decreases significantly
with increasing averaging.

The relationships and density distributions that we determine nicely
demonstrate that most of the data points in the quiet-Sun region are
characterized by low magnetic flux and low chromospheric emission.
In other words, the majority of data points lies in the lower left
quarter of the scatter plot and ``anchors'' the relation. There are
too few data points with high chromospheric emission and enhanced
magnetic flux (greater than 100~G) to very reliably determine the
form of the relation (see Figs.~\ref{fig3} and \ref{fig4}). The
rather small $1\sigma$ error bars of the fitted exponents (last
column of Table~2) are the formal errors returned by the fitting
procedure and appear rather optimistic to us.

\begin{figure}
   \centering
   \includegraphics[width=0.35\textwidth]{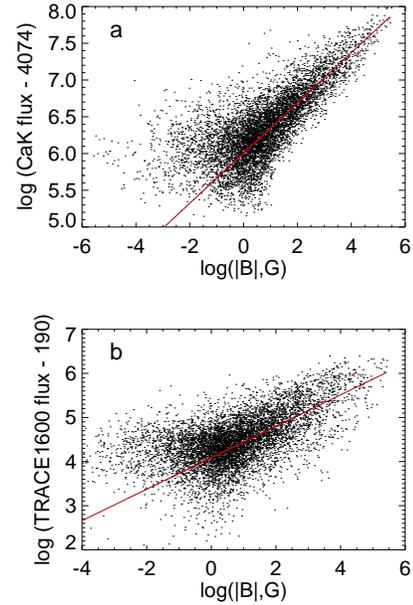}
   \caption{Log-log scatter plots for \textbf{a)} the \ca-line excess mean intensity
   (after the subtraction of the zero flux) and \textbf{b)} TRACE 1600
   \AA\ excess mean intensity vs. the absolute value of
   the magnetic flux density (for 2\arcsec\ pixel size).
   The solid lines represent the least-squares power-law fits to the data.}
              \label{fig6}%
    \end{figure}

Note that fitting of the data directly with a power law containing a
zero flux item $I_{0}$, corresponding to emission from locations of
zero magnetic flux (non-magnetic component) is similar to
determining the zero point of chromospheric emission using the
algorithm applied by \citet{1989ApJ...337..964S}, according to which
a range of test values of $I_{0}$ is subtracted from the observed
flux $I$ until the mean relationship $\Delta
I=I-I_{0}=a\cdot|B|^{b}$ reaches its highest correlation by
minimizing the perpendicular offsets between the excess flux and the
fit. We have tried fitted the data using both algorithms and obtained
similar zero flux values and power law exponents. To demonstrate the
success of the algorithm, introduced by \citet{1989ApJ...337..964S},
applied to the data analyzed here, we plot in Fig.~\ref{fig6} the
excess \ca-line and 1600~\AA\ fluxes after subtraction of the
derived basal flux (listed in the Y-axis captions) and the
corresponding power-law curves. These fits yield zero-flux
values $I_{0}$ and power law exponents $b$ very similar to those
derived when fitting the observed fluxes directly with a function of the form
$I=I_{0}+a\cdot|B|^{b}$ (listed in
Table~\ref{table2}).

The plot of millimeter brightness vs. magnetic field strength displays a
considerable scatter about the mean relation, which may have
different causes (see Fig.~\ref{fig5}). Pixels in the lower left
corner with very low mm brightness can result from the
underestimation of the mm flux due to the high noise in the snapshot
images (White et al. 2006). Some of the peculiar pixels with intense
mm brightness (in the upper left corner) are located above localized
(small-scale) photospheric magnetic elements and near polarity
changes. Therefore the corresponding magnetic flux almost vanishes
when degrading the magnetogram to the BIMA resolution. Again, a
power law gives a far better representation of the data than the
linear curve. In this case, a rather low exponent of $0.20$ is
obtained.

\subsection{Magnetogram signal thresholding experiment}
\begin{figure}
   \centering
   \includegraphics[width=0.2\textwidth,angle=90]{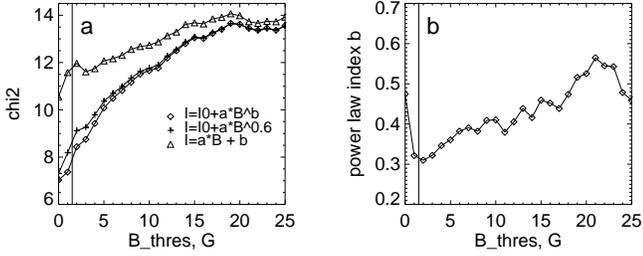}
   \caption{Thresholding experiment for the CaII K-line flux vs. absolute value of the magnetic field:
     \textbf{a)} $\chi^2$ and \textbf{b)} 
     power law exponent 
     as a function of the imposed B threshold.
   The vertical line represents the $1\sigma$-noise level of the magnetogram signal.}
              \label{fig7}%
    \end{figure}

\begin{figure}
   \centering
   \includegraphics[width=0.2\textwidth,angle=90]{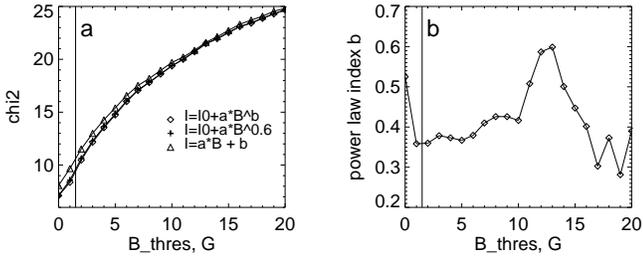}
   \caption{The same as in Fig.~\ref{fig6} for the TRACE 1600~\AA\ flux.}
              \label{fig8}%
    \end{figure}

In Figs.~\ref{fig3}, \ref{fig4} and \ref{fig5} the brightness shows
signs of increasing more rapidly for smaller fields with $B \leq
30$~G. This may be saying that the chromospheric emission from
weak-field regions (internetwork) has a different source than from
stronger network fields. To study how the correlation depends on the
magnetogram signal we have performed a thresholding experiment. We
have successively fitted the relations between the chromospheric
radiation and photospheric magnetic flux exceeding a number of
threshold values (from $0$ to $35$ Gauss for correlation at the
resolution of the photospheric magnetogram and from $0$ to $20$
Gauss for the resolution of the mm images). In Figs.~\ref{fig7} and
\ref{fig8} we compare the resuting $\chi^2$ values for linear and
power law fits as a function of the $B$ threshold value, plotted with
triangle and diamond symbols, respectively. We also show the
dependence of the fitted power-law exponent on the threshold value.

The power-law exponent of the fit to the calcium intensity grows
with the imposed $B$ threshold (see Fig.~\ref{fig7}). The value of
the power-law exponent of $\simeq$ 0.5 is reached both when we discard
pixels with weak fields and consider only those pixels with magnetic
flux exceeding $\simeq 20$~G, and when we include all pixels in the fit
(including those dominated by noise). The dependence of the power-law index on B
threshold for the intensity at 1600~\AA\ shows a different
behaviour. It increases until it reaches a value of about $0.6$. For
magnetic flux density exceeding $13$~G it starts to decrease rapidly
for higher thresholds (Fig.~\ref{fig8}b). The scatter of the points
is so large that for these data it is not possible to clearly
distinguish between linear and power-law fits. The sensitivity of
the power law parameters to the $B$ threshold is due to the small
number of points at large $<B>$ (see Fig.~\ref{fig4}), which is due
to the fact that the observations refer to the quiet Sun.

The scatter in the intensity at 3.5 mm is so large that its
relationship with the magnetic flux density can be equally well fit
by a linear or a power-law function. Also, the $\chi^2$ values of
the fits are large.
However, according to the thresholding experiment for the
correlation between the mm intensity and the magnetic flux (not
plotted), a power law adequately fits the data for magnetic fields
stronger than 3~G ($\sim 1\sigma$-noise level) and weaker than 15~G.
For weak fields, the large scatter affects significantly the quality of
the fit, while for fields stronger than 15~G there are insufficient
data pixels for confident power-law fitting.

 \begin{figure}
   \centering
   \includegraphics[width=0.27\textwidth]{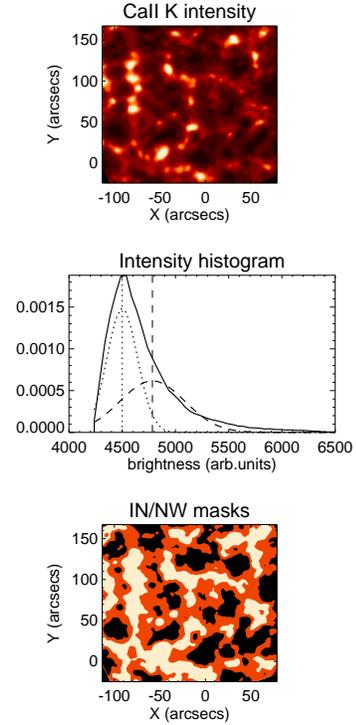}
   \caption{IN/NW definition. Upper panel: CaII K-line mean intensity image. Central panel:
   mean intensity histogram (thick
   solid), single-Gaussian profiles (dotted, dashed) from a double-Gaussian fit to the data points
   with vertical lines of the same type marking
   the location of the Gaussian maxima. Bottom panel: Masks used for the definition of the
   internetwork (black) and network (white) areas.}
              \label{fig10}%
    \end{figure}

\subsection{Discrimination between IN and NW emission. Separate flux analysis}

Next we study the relations between the chromospheric emission and
the magnetic flux separately for the areas representing cell
interiors or internetwork (IN) and chromospheric network (NW).
In other words, we tried to separate magnetically weak-field
(IN locations) and strong-field (NW locations) components of the
emission. To discriminate between IN and NW locations we employed a
brightness threshold technique \citep[see,
e.g.][]{2001A&A...379.1052K}. To this end we fitted the \ca-line
intensity histogram constructed for the time-averaged intensity map
by a double-Gaussian function, assuming that the intensity histogram
is dominated by two main populations representing the IN and NW intensities
(see Fig.~\ref{fig10}).
The resulting maxima of the single-Gaussian components were used as
intensity thresholds to define respectively internetwork and network
areas on the mean map. The resulting IN (black) and NW (white) binary
masks are shown in the bottom panel of Fig.~\ref{fig10}. The points
belonging to neither of these categories are indicated by an
intermediate shading.

\begin{figure}
   \centering
   \includegraphics[width=0.35\textwidth,angle=90]{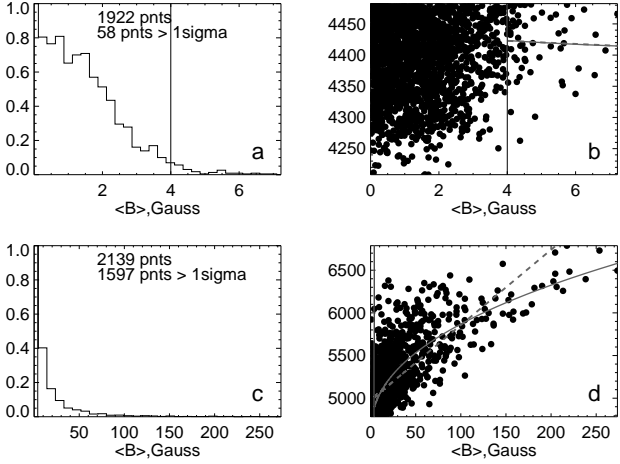}
   \caption{Comparison of IN and NW emission averaged over 0.5h:  histograms of \textbf{a)} IN and \textbf{c)} NW magnetogram signal,
   \textbf{b, d)} \ca-line flux vs. absolute value of the IN and NW magnetic field, respectively.
   The solid vertical lines represent $1\sigma$-noise of the magnetogram signal.}
              \label{fig11}%
    \end{figure}

\begin{figure}
   \centering
   \includegraphics[width=0.35\textwidth,angle=90]{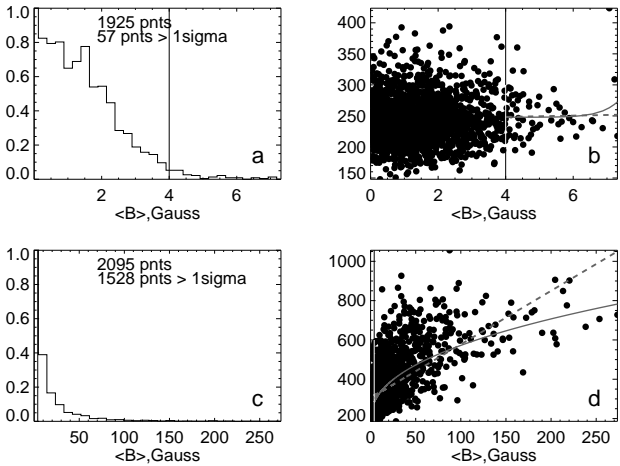}
   \caption{The same as in Fig.~\ref{fig11} for 1600~\AA\ flux.}
              \label{fig12}%
    \end{figure}

\begin{figure}
   \centering
   \includegraphics[width=0.35\textwidth,angle=90]{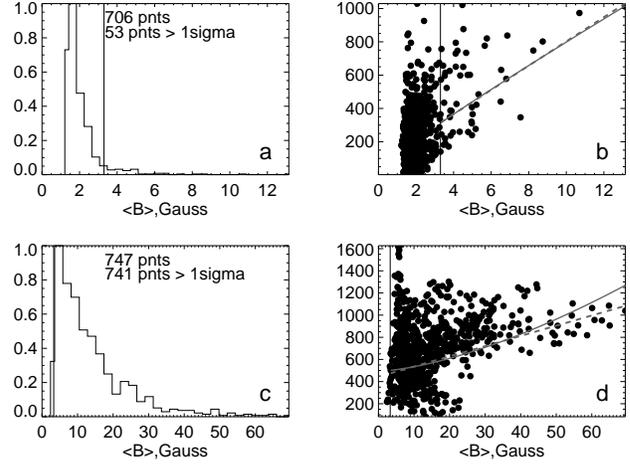}
   \caption{The same as in Fig.~\ref{fig11} for 3.5~mm brightness.}
              \label{fig14}%
    \end{figure}

We apply these binary masks to study the
relationship between the photospheric magnetic field and IN/NW
fluxes, respectively (see Figs.~\ref{fig11},\ref{fig12} and
\ref{fig14}). Following the approach outlined in Sect.~4.3 we fitted
the scatter plots with linear and power-law functions. For the areas
selected to represent the network a power law provides the best fit
to the data. In Table~\ref{table3} we list the exponents from the
best-fit power laws for the images at the resolution of the MDI
magnetograms (4\arcsec) and of the BIMA images (12\arcsec). Below we
discuss the results obtained for the 3.5h image averages, but for
comparison we list in Table~\ref{table3} also those for the 0.5h
time-averages. For the calcium NW intensity, averaged in time over 3.5
hours, the best-fit power-law exponent is 0.6, which is consistent
with the value derived by \citet{1989ApJ...337..964S} for the
\ca-line flux in an active region and by \citet{1999ApJ...515..812H}
for the enhanced network. We note a big difference from the value we
derived with no image segmentation (0.6 vs. 0.31, see
Table~\ref{table2}), which means a stronger dependence of the
network calcium intensity on the magnetic flux. At a resolution of
12\arcsec\ the best fit for the NW \ca-line flux is also provided by
a power law. This time the power law exponent is 0.54, which is also
in agreement with the results of \citet{1989ApJ...337..964S}, who
state that the mean dependence of the \ca-line excess flux on the
magnetic flux density is more or less scale independent. Analysis of
the network emission at 1600~\AA\ gives similar results. The
power-law exponent in the relationship between the NW intensity at
1600~\AA\ and the magnetic flux density is $\simeq0.5$ at 4\arcsec\
resolution and $0.66$ at 12\arcsec\ resolution (see
Table~\ref{table3}). Finally, millimeter emission from the network
regions as a function of the absolute value of the magnetic field is
also best described by a power law (among the considered functions,
see Fig.~\ref{fig14}). In this case, however, exponents larger than
unity are obtained and we regard this result with caution and do not
include it in Table~\ref{table3}.

\medskip

\medskip

\begin{table}
  \centering
  \caption{Dependence of the network emission on the magnetic flux: exponents of the power-law fits calculated
  for the images at 4\arcsec\ and 12\arcsec\ resolution, averaged over 3.5h and 0.5h}\label{table3}
    \begin{tabular}{ccccc}
            \hline\hline
            \noalign{\smallskip}
            NW flux & \multicolumn{2}{c}{4\arcsec\ resolution} & \multicolumn{2}{c}{12\arcsec\ resolution} \\
             & 3.5h & 0.5h & 3.5h & 0.5h \\
            \noalign{\smallskip}
            \hline
            \noalign{\smallskip}
            $|B| - I_{CaII K}$ & 0.6 & 0.43 & 0.54 & 0.47   \\
             \noalign{\smallskip}
            $|B| - I_{1600 \AA}$ & 0.49 & 0.53 & 0.66 & 0.44  \\
            \noalign{\smallskip}
            \hline
    \end{tabular}

\end{table}

The fact that chromospheric emission (be it from \ca, the 1600\AA\
range or 3.5 mm continuum) from the internetwork regions shows
almost no dependence on the magnetogram signal (see upper panels of
Figs.~\ref{fig11}, \ref{fig12} and \ref{fig14}) is not really
surprising given that the magnetogram values in the internetwork do
not exceed 10 Gauss, with mean of 1.5 Gauss and 1.0 Gauss (for 0.5h
and 3.5h time averages, respectively). Therefore, even after
creation of 0.5h averaged images, which reduces significantly the
noise in the magnetograms, only about 3\% of IN points display $B$
values exceeding the 1$\sigma$ noise level (see left panels of
Figs.~\ref{fig11} and \ref{fig12}). Even after averaging over 3.5
hours, only 20\% of IN locations have $B$ values greater than
1$\sigma$-noise. Data that have been so heavily averaged are only of
limited value to study the internetwork as IN features are commonly
characterized by much shorter time scales. We also studied the
internetwork emission using a less strict criterion to discriminate
between the cell interiors and the cell boundaries. Points with the
calcium intensity not exceeding the value corresponding to the
intercept point of the two-Gaussian fit to the intensity histogram
were selected to represent the IN mask. However, the use of this
mask does not influence the results of the fit. Again, no dependence
of IN chromospheric emission on the magnetogram signal was found.
Consequently, the high level of noise in the MDI magnetograms used
here (compared to the range of the $B$ values in the internetwork
regions) hinders a reliable study of the relation of internetwork
chromospheric emission on magnetic field.

\begin{figure}
   \centering
   \includegraphics[width=0.35\textwidth]{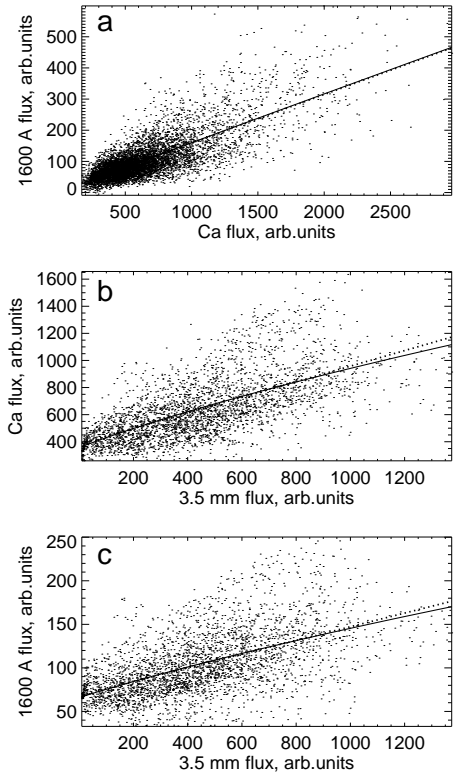}
   \caption{Scatter plots of chromospheric fluxes and their power-law fits (plotted as solid lines).
   For comparison we also plot linear fits as dotted lines. \textbf{a)}
   1600~\AA\ excess flux vs. CaII K-line excess flux, \textbf{b)} CaII
K-line excess flux vs. 3.5~mm flux, and \textbf{c)} 1600~\AA\ excess
flux vs. 3.5~mm flux.}
              \label{fig13}%
    \end{figure}

\subsection{Chromospheric flux-flux relations}

We studied flux-flux relationships by constructing scatter plots for
chromospheric intensities. In this analysis excess fluxes for K-line
and 1600~\AA\ emission were considered, employing the basal fluxes
derived earlier (see Sect.~4.3 and Fig.~\ref{fig6}). For the mm
intensity we included the zero flux as a parameter in the fitting
power-law function. Next, using the $\chi^2$ criterion we searched
for the best linear and power law fits to these relationships. The
scatter plots of intensity at 1600~\AA\ vs. \ca-line intensity as
well as \ca-line and 1600~\AA\ vs. 3.5~mm intensity are depicted in
Fig.~\ref{fig13} together with the best power-law and linear fits.
For all three dependencies the curves depicting both fits are hardly
distinguishable and the fits are of similar quality (see
Fig.~\ref{fig13}).

For the relation between K-line intensity and intensity at 1600~\AA\
the power-law exponent is $\simeq1$, which implies that the
relationship between the excess fluxes is close to linear at a
resolution of 4\arcsec. The relationship between mm and calcium, as
well as between mm and UV emissions at the resolution of the mm
images, are both best described by power laws with the best-fit
exponents of $\simeq0.7$, but as mentioned this exponent is rather
uncertain since a linear function also represents the data well.

\section{DISCUSSION AND CONCLUSIONS}

We have considered chromospheric disk-center, quiet-Sun emission at
different wavelengths and its dependence on photospheric magnetogram signal,
including, for the first time, observations at 3.5~mm. Maps in \ca\
core intensity, the TRACE 1600~\AA\ channel, BIMA 3.5 mm brightness
temperature and (unsigned) MDI magnetograms reveal clearly the same
pattern, although at different spatial
resolutions. At the resolution of 12\arcsec\ millimeter brightness
resembles quiet-Sun time-averaged emission from other chromospheric
heights, and demonstrates a dependence on the underlying
magnetic field similar to that of the other chromospheric
diagnostics.

The tightest relation between the chromospheric emission and
photospheric magnetic field is displayed by the \ca\ line. At all
wavelengths a power law gives a lower $\chi^2$ than a linear fit,
although for the 1600~\AA\ radiation the difference is marginal. For
\ca\ this result supports the conclusion reached by
\citet{1989ApJ...337..964S}, but based on active region data, and by
\citet{1999ApJ...515..812H} for both active and quiet-Sun magnetic
structures. For \ca\ \citet{1975ApJ...200..747S} and
\citet{1998SoPh..179..253N} found a linear dependence to be
adequate. However, the linear dependencies were obtained after
excluding weak fields (weaker than 25~G and 20~G, respectively) and
for a magnetogram signal not exceeding 120~G and 50~G, respectively.
In our study we included both weak fields exceeding $1\sigma$-noise
level of the magnetograms (1.5 or 4~G depending on the amount of
averaging), and the highest values reached in the quiet-Sun data 
(about 230~G at MDI's spatial resolution). This suggests that the
non-linearity of the relation found in this work is influenced
mostly by weak chromospheric emission that reveals a strong
dependence on magnetic fields. For instance, if we set a $B$
threshold of 20~G we cannot distinguish between the quality of the
linear and power-law fits for all chromospheric emissions analyzed
here. Our conclusions agree with those presented by
\citet{2007A&A...466.1131R}, who found from the POLIS measurements
at 1\arcsec\ resolution a clear non-linear increase of the calcium
core flux with magnetic flux for values $B \leq 100$~G and a slow
increase for higher values. They also demonstrated that the
power-law exponent depends strongly on the value of the lower
threshold.

The difference between the maximum magnetogram signal reached in our
data and that in the data of \citet{1975ApJ...200..747S} and
\citet{1998SoPh..179..253N} may indicate either that their quiet-Sun
regions are even quieter than the one we analyzed, or it may reflect
differences in spatial resolution.
One more difference between our analysis and the others is that we
consider temporally averaged data rather than snapshot images. By
using time-averaging we reduce significantly the scatter in the
relationship and the noise in the magnetograms, which lead to more
reliable power-law fits. By employing snapshot images we get more
points in the scatter plots but due to higher magnetogram noise
level the power-law exponent is even lower (and less reliable) than
for the time-averaged images. One should also note the difference in
the analysis techniques between this work and the others applied for
building the scatter plots. In this work we employed pixel-to-pixel
comparison of the images. However, the results of the fits are
similar if we consider the binned chromospheric data averaged within
a narrow range of magnetogram values. On the whole, from the data
analyzed here we are able to confidently determine non-linear
behaviour of 3 chromospheric magnetic-intensity relationships.

The exact power law exponents of the relations between the calcium
core emission and the photospheric magnetic field differs from that
found by \citet{1989ApJ...337..964S} (0.3 vs. 0.6) . This probably
has to do with the smaller $<B>$ range available in the quiet Sun,
although it cannot be ruled out that the \ca\ intensity does not
increase so rapidly in the quiet Sun as in active regions
\citep[see][]{1991A&A...250..220S}. To compare these explanations,
we studied the dependence of the calcium emission on the magnetic
flux in an active region close to disk center observed by the
BBSO and MDI on 11.03.2004. The exponent obtained from a power-law
fit for the AR emission is surprisingly low and similar to the QS
values, being in the range 0.28-0.32, with the exact value depending
somewhat on the criterion chosen to identify the AR pixels on the map. The
use of a pixel size and a noise threshold similar to that of
\citet{1989ApJ...337..964S} does not increase the value of the
power-law exponent and it still differs from that of
\citet{1989ApJ...337..964S} and \citet{1999ApJ...515..812H}. The
difference may have to do with the spectral range covered by the
different observations.

The reduction in the power-law exponent as compared to the
value derived by \citet{1989ApJ...337..964S} may be to a large extent
the result of the inclusion of weak magnetic flux regions, which are
systematically characterized by magnetogram field strengths that are
smaller than the true values due to the presence of weak neutral lines with null
line-of-sight component. Unfortunately, we cannot exclude the
influence of these horizontal magnetic fields, which are missed by
MDI observations. Such fields, partly seen in the internetwork
regions by Hinode \citep[e.g.][]{2007PASJ...59S.571L}, are very
likely small loops
\citep[][]{2007ApJ...666L.137C,2007A&A...469L..39M} and may be due
to a local dynamo
\citep[][]{2007A&A...465L..43V,2008A&A...481L...5S}. It is unclear
if such low-lying loops really influence the structure of the
higher--lying
chromosphere. However, if they do contribute to the heating of those
layers then this could explain the high chromospheric emission from
many spatial locations associated with small magnetic fluxes in the
MDI data (Fig.~\ref{fig6}). Unresolved opposite magnetic polarities
within the same individual MDI pixels have a similar effect. Both these
effects together could explain the result that the power law exponent is lower
when very quiet (low flux) pixels are included. Studies at higher
spatial resolution and sensitivity are needed to test these
hypotheses.

On the whole, chromospheric emission displays a rather different
dependence on the network and internetwork magnetic field. The
power-law exponents in the relation between the \ca-line intensity
and the network magnetic flux are found to be similar to those
obtained by \citet{1989ApJ...337..964S} for an active region and by
\citet{1999ApJ...515..812H} for the quiet Sun network.
The same conclusion holds for the relation between the UV
emission at 1600~\AA\ and magnetic flux in the network regions of
the quiet Sun. For chromospheric fluxes at a resolution of 4\arcsec\
it was found that emission from the network entirely defines the
trend for magnetic flux densities stronger than $\simeq 20$~G. Our
results for the calcium core emission remarkably confirm those of
\citet{2007A&A...466.1131R}, who obtained for the ensemble of
quiet-Sun points a low value for the power-law exponent of
$\simeq$0.2 (for snapshot images), while for the network emission
from regions with $B \geq 20$ G the exponent reaches the value
of 0.51 (and 0.45 for $B \geq 10$ G).

A considerable scatter about the mean relation between the
millimeter flux at 3.5~mm and the magnetic flux density per
12\arcsec\ aperture makes a confident study of the mean trend
difficult. Although the data strongly suggests that the dependence
of the 3.5~mm brightness on magnetic field is non-linear and best
described by a power law (among the functional forms considered here), but the
sizeable scatter does not allow the exponents to be determined
reliably. The best fit exponent depends strongly on the
range of $<B>$ considered.

In the internetwork regions almost no dependence of the \ca-line, UV
at 1600~\AA\ and 3.5~mm intensity on the magnetogram
signal was found.
Unfortunately, due to the very low $B$ values in the internetwork,
only a few percent of IN structures are associated with magnetogram
signal exceeding the noise level. However, it should be noted that analysis of
the cell interiors and their relation to the underlying magnetic
field carried out in \citet{2007A&A...466.1131R} for \ion{Ca}{ii} H-line
with lower noise and higher resolution has revealed similar results.
This supports the idea that the role of magnetic field is
negligible for bright internetwork grains (e.g.
acoustic heating).

Correlation analysis performed for the quiet-Sun chromospheric
emissions surprisingly revealed a linear relation between the
\ca-line intensity and UV brightness at 1600~\AA. According to
\citet{1991A&A...252..203R}, who claimed that the relations between
the fluxes originating in the same atmospheric regimes are commonly
close to linear, these chromospheric emissions are both
generated in the chromospheric temperature regime from chromospheric
ranges of heights. However, solar models place UV emission at
1600~\AA\ in the temperature minimum region, while calcium emission
is believed to be generated from heights more typical of the lower
chromosphere. The power-law relations between calcium flux and
3.5~mm flux, as well as between 3.5~mm brightness and 1600~\AA\
flux, suggest different temperatures of formation for these pairs of
emissions, but as the power-law exponent do not differ much from
unity we may assume that these emissions are still close in the
heights of formation. However, the flux correlation analysis
for the considered emissions and its conclusions should be treated
with caution. Due to the very different temperature sensitivity of,
on the one hand, UV and calcium intensity, and, on the other hand,
3.5 mm brightness, they represent different contributions from small
(unresolved) hot magnetic features and cool weak-field regions
within the same resolution element. In other words, in an
inhomogeneous atmosphere the UV samples mainly the hot features,
while the mm is composed of radiation from both. This could explain
some of the difference in the behaviour of 1600~\AA\ and \ca\ excess
emission compared to mm emission, and, in particular, the large
scatter in the flux-flux plots (Fig.~\ref{fig13}).

Summarizing, clear spatial correlations found
between photospheric magnetograms, the \ca-line, UV intensity at
1600~\AA\ and 3.5~mm emission indicate that heating in the quiet-Sun
lower and middle chromosphere maps out the underlying photospheric
magnetic field rather well. Our results imply that at the
chromospheric heights covered by the wavelengths investigated here we
deal with the same heating mechanism. However, any mechanism
that explains chromospheric heating must explain the relationship
between chromospheric intensity excess and magnetic fields. In
particular, it should validate both the dominating role of magnetic
field in the chromospheric emission of the network and the absence
of correlation between the internetwork emission and magnetic flux.

Of particular interest is the possibility of studying the middle
chromosphere at the heights where mm emission originates, but
because of very limited spatial resolution and insufficient
sensitivity a detailed study like that cannot be conducted at
present. In general, present data on both chromospheric emission and
magnetic field are not sufficient to establish physical mechanisms
that are acting at chromospheric heights leading to the observed
brightness structures. In this respect, an extraordinarily powerful
tool to study the thermal structure of the solar atmosphere from the
temperature minimum region to the middle chromosphere will be
provided by the Combined Array for Research in Millimeter--wave
Astronomy (CARMA) and particularly by Atacama Large Millimeter Array
(ALMA) as was demonstrated in \citet{2008Ap&SS.313..197L}.

\acknowledgements We thank H. Wang and V. Yurchyshyn for providing
us with the data from the BBSO. We gratefully acknowledge the
open-data policies of the NASA satellite TRACE and the ESA/NASA
satellite SOHO and their instrument teams. We thank the referee,
Charlie Lindsey, for his considerate and helpful comments.
Millimeter--wavelength astronomy at the University of Maryland is
supported by NSF grant AST--0540450. Solar research at the
University of Maryland is supported by NASA grants NNX06--AC18G and
NNX08--AQ48G. M. Loukitcheva acknowledges support from the Grant
MK-1853.2007.2 of the Grant Council of the President of the Russian
Federation.

\bibliographystyle{aa} 
\bibliography{1133} 

\begin{thebibliography}{37}
\expandafter\ifx\csname natexlab\endcsname\relax\def\natexlab#1{#1}\fi

\bibitem[{{Brandt} {et~al.}(1992){Brandt}, {Rutten}, {Shine}, \&
  {Trujillobueno}}]{1992ASPC...26..161B}
{Brandt}, P.~N., {Rutten}, R.~J., {Shine}, R.~A., \& {Trujillobueno}, J. 1992,
  in Astronomical Society of the Pacific Conference Series, Vol.~26, Cool
  Stars, Stellar Systems, and the Sun, ed. M.~S. {Giampapa} \& J.~A.
  {Bookbinder}, 161

\bibitem[{{Centeno} {et~al.}(2007){Centeno}, {Socas-Navarro}, {Lites}, {Kubo},
  {Frank}, {Shine}, {Tarbell}, {Title}, {Ichimoto}, {Tsuneta}, {Katsukawa},
  {Suematsu}, {Shimizu}, \& {Nagata}}]{2007ApJ...666L.137C}
{Centeno}, R., {Socas-Navarro}, H., {Lites}, B., {et~al.} 2007, \apjl, 666,
  L137

\bibitem[{{Denker} {et~al.}(2001){Denker}, {Yang}, \&
  {Wang}}]{2001SoPh..202...63D}
{Denker}, C., {Yang}, G., \& {Wang}, H. 2001, \solphys, 202, 63

\bibitem[{{Fontenla} {et~al.}(1993){Fontenla}, {Avrett}, \&
  {Loeser}}]{1993ApJ...406..319F}
{Fontenla}, J.~M., {Avrett}, E.~H., \& {Loeser}, R. 1993, \apj, 406, 319

\bibitem[{{Fontenla} {et~al.}(2007){Fontenla}, {Curdt}, {Avrett}, \&
  {Harder}}]{2007A&A...468..695F}
{Fontenla}, J.~M., {Curdt}, W., {Avrett}, E.~H., \& {Harder}, J. 2007, \aap,
  468, 695

\bibitem[{{Fossum} \& {Carlsson}(2005)}]{2005ApJ...625..556F}
{Fossum}, A. \& {Carlsson}, M. 2005, \apj, 625, 556

\bibitem[{{Handy} {et~al.}(1999){Handy}, {Acton}, {Kankelborg}, {Wolfson},
  {Akin}, {Bruner}, {Caravalho}, {Catura}, {Chevalier}, {Duncan}, {Edwards},
  {Feinstein}, {Freeland}, {Friedlaender}, {Hoffmann}, {Hurlburt}, {Jurcevich},
  {Katz}, {Kelly}, {Lemen}, {Levay}, {Lindgren}, {Mathur}, {Meyer}, {Morrison},
  {Morrison}, {Nightingale}, {Pope}, {Rehse}, {Schrijver}, {Shine}, {Shing},
  {Strong}, {Tarbell}, {Title}, {Torgerson}, {Golub}, {Bookbinder}, {Caldwell},
  {Cheimets}, {Davis}, {Deluca}, {McMullen}, {Warren}, {Amato}, {Fisher},
  {Maldonado}, \& {Parkinson}}]{1999SoPh..187..229H}
{Handy}, B.~N., {Acton}, L.~W., {Kankelborg}, C.~C., {et~al.} 1999, \solphys,
  187, 229

\bibitem[{{Handy} {et~al.}(1998){Handy}, {Bruner}, {Tarbell}, {Title},
  {Wolfson}, {Laforge}, \& {Oliver}}]{1998SoPh..183...29H}
{Handy}, B.~N., {Bruner}, M.~E., {Tarbell}, T.~D., {et~al.} 1998, \solphys,
  183, 29

\bibitem[{{Harvey} \& {White}(1999)}]{1999ApJ...515..812H}
{Harvey}, K.~L. \& {White}, O.~R. 1999, \apj, 515, 812

\bibitem[{{Krijger} {et~al.}(2001){Krijger}, {Rutten}, {Lites}, {Straus},
  {Shine}, \& {Tarbell}}]{2001A&A...379.1052K}
{Krijger}, J.~M., {Rutten}, R.~J., {Lites}, B.~W., {et~al.} 2001, \aap, 379,
  1052

\bibitem[{{Leighton}(1959)}]{1959ApJ...130..366L}
{Leighton}, R.~B. 1959, \apj, 130, 366

\bibitem[{{Lindsey} {et~al.}(1995){Lindsey}, {Kopp}, {Clark}, \&
  {Watt}}]{1995ApJ...453..511L}
{Lindsey}, C., {Kopp}, G., {Clark}, T.~A., \& {Watt}, G. 1995, \apj, 453, 511

\bibitem[{{Lindsey} \& {Jefferies}(1991)}]{1991ApJ...383..443L}
{Lindsey}, C.~A. \& {Jefferies}, J.~T. 1991, \apj, 383, 443

\bibitem[{{Lites} {et~al.}(2007){Lites}, {Socas-Navarro}, {Kubo}, {Berger},
  {Frank}, {Shine}, {Tarbell}, {Title}, {Ichimoto}, {Katsukawa}, {Tsuneta},
  {Suematsu}, {Shimizu}, \& {Nagata}}]{2007PASJ...59S.571L}
{Lites}, B., {Socas-Navarro}, H., {Kubo}, M., {et~al.} 2007, \pasj, 59, 571

\bibitem[{{Lites} {et~al.}(1999){Lites}, {Rutten}, \&
  {Berger}}]{1999ApJ...517.1013L}
{Lites}, B.~W., {Rutten}, R.~J., \& {Berger}, T.~E. 1999, \apj, 517, 1013

\bibitem[{{Loukitcheva} {et~al.}(2008){Loukitcheva}, {Solanki}, \&
  {White}}]{2008Ap&SS.313..197L}
{Loukitcheva}, M.~A., {Solanki}, S.~K., \& {White}, S. 2008, \apss, 313, 197

\bibitem[{{Mart{\'{\i}}nez Gonz{\'a}lez} {et~al.}(2007){Mart{\'{\i}}nez
  Gonz{\'a}lez}, {Collados}, {Ruiz Cobo}, \& {Solanki}}]{2007A&A...469L..39M}
{Mart{\'{\i}}nez Gonz{\'a}lez}, M.~J., {Collados}, M., {Ruiz Cobo}, B., \&
  {Solanki}, S.~K. 2007, \aap, 469, L39

\bibitem[{{Nindos} \& {Zirin}(1998)}]{1998SoPh..179..253N}
{Nindos}, A. \& {Zirin}, H. 1998, \solphys, 179, 253

\bibitem[{{Ortiz} \& {Rast}(2005)}]{2005MmSAI..76.1018O}
{Ortiz}, A. \& {Rast}, M. 2005, Memorie della Societa Astronomica Italiana, 76,
  1018

\bibitem[{{Pauluhn} {et~al.}(2000){Pauluhn}, {Solanki}, {R{\"u}edi}, {Landi},
  \& {Sch{\"u}hle}}]{2000A&A...362..737P}
{Pauluhn}, A., {Solanki}, S.~K., {R{\"u}edi}, I., {Landi}, E., \&
  {Sch{\"u}hle}, U. 2000, \aap, 362, 737

\bibitem[{{Remling} {et~al.}(1996){Remling}, {Deubner}, \&
  {Steffens}}]{1996A&A...316..196R}
{Remling}, B., {Deubner}, F.-L., \& {Steffens}, S. 1996, \aap, 316, 196

\bibitem[{{Rezaei} {et~al.}(2007){Rezaei}, {Schlichenmaier}, {Beck}, {Bruls},
  \& {Schmidt}}]{2007A&A...466.1131R}
{Rezaei}, R., {Schlichenmaier}, R., {Beck}, C.~A.~R., {Bruls}, J.~H.~M.~J., \&
  {Schmidt}, W. 2007, \aap, 466, 1131

\bibitem[{{Rutten} {et~al.}(1991){Rutten}, {Schrijver}, {Lemmens}, \&
  {Zwaan}}]{1991A&A...252..203R}
{Rutten}, R.~G.~M., {Schrijver}, C.~J., {Lemmens}, A.~F.~P., \& {Zwaan}, C.
  1991, \aap, 252, 203

\bibitem[{{Rutten} {et~al.}(1999){Rutten}, {de Pontieu}, \&
  {Lites}}]{1999ASPC..183..383R}
{Rutten}, R.~J., {de Pontieu}, B., \& {Lites}, B. 1999, in Astronomical Society
  of the Pacific Conference Series, Vol. 183, High Resolution Solar Physics:
  Theory, Observations, and Techniques, ed. T.~R. {Rimmele}, K.~S.
  {Balasubramaniam}, \& R.~R. {Radick}, 383

\bibitem[{{Rutten} \& {Uitenbroek}(1991)}]{1991SoPh..134...15R}
{Rutten}, R.~J. \& {Uitenbroek}, H. 1991, \solphys, 134, 15

\bibitem[{{Scherrer} {et~al.}(1995){Scherrer}, {Bogart}, {Bush}, {Hoeksema},
  {Kosovichev}, {Schou}, {Rosenberg}, {Springer}, {Tarbell}, {Title},
  {Wolfson}, {Zayer}, \& {MDI Engineering Team}}]{1995SoPh..162..129S}
{Scherrer}, P.~H., {Bogart}, R.~S., {Bush}, R.~I., {et~al.} 1995, \solphys,
  162, 129

\bibitem[{{Schrijver}(1987)}]{1987A&A...172..111S}
{Schrijver}, C.~J. 1987, \aap, 172, 111

\bibitem[{{Schrijver}(1992)}]{1992A&A...258..507S}
{Schrijver}, C.~J. 1992, \aap, 258, 507

\bibitem[{{Schrijver} {et~al.}(1989){Schrijver}, {Cote}, {Zwaan}, \&
  {Saar}}]{1989ApJ...337..964S}
{Schrijver}, C.~J., {Cote}, J., {Zwaan}, C., \& {Saar}, S.~H. 1989, \apj, 337,
  964

\bibitem[{{Sch{\"u}ssler} \& {V{\"o}gler}(2008)}]{2008A&A...481L...5S}
{Sch{\"u}ssler}, M. \& {V{\"o}gler}, A. 2008, \aap, 481, L5

\bibitem[{{Sivaraman} {et~al.}(2000){Sivaraman}, {Gupta}, {Livingston},
  {Dam{\'e}}, {Kalkofen}, {Keller}, {Smartt}, \& {Hasan}}]{2000A&A...363..279S}
{Sivaraman}, K.~R., {Gupta}, S.~S., {Livingston}, W.~C., {et~al.} 2000, \aap,
  363, 279

\bibitem[{{Sivaraman} \& {Livingston}(1982)}]{1982SoPh...80..227S}
{Sivaraman}, K.~R. \& {Livingston}, W.~C. 1982, \solphys, 80, 227

\bibitem[{{Skumanich} {et~al.}(1975){Skumanich}, {Smythe}, \&
  {Frazier}}]{1975ApJ...200..747S}
{Skumanich}, A., {Smythe}, C., \& {Frazier}, E.~N. 1975, \apj, 200, 747

\bibitem[{{Solanki}(2004)}]{2004IAUS..223..195S}
{Solanki}, S.~K. 2004, in IAU Symposium, Vol. 223, Multi-Wavelength
  Investigations of Solar Activity, ed. A.~V. {Stepanov}, E.~E.
  {Benevolenskaya}, \& A.~G. {Kosovichev}, 195--202

\bibitem[{{Solanki} {et~al.}(1991){Solanki}, {Steiner}, \&
  {Uitenbroeck}}]{1991A&A...250..220S}
{Solanki}, S.~K., {Steiner}, O., \& {Uitenbroeck}, H. 1991, \aap, 250, 220

\bibitem[{{V{\"o}gler} \& {Sch{\"u}ssler}(2007)}]{2007A&A...465L..43V}
{V{\"o}gler}, A. \& {Sch{\"u}ssler}, M. 2007, \aap, 465, L43

\bibitem[{{White} {et~al.}(2006){White}, {Loukitcheva}, \&
  {Solanki}}]{2006A&A...456..697W}
{White}, S.~M., {Loukitcheva}, M., \& {Solanki}, S.~K. 2006, \aap, 456, 697

\end{thebibliography}

\end{document}